\documentclass[aps,pra,superscriptaddress,twocolumn,nofootinbib, nobibnotes]{revtex4}
\usepackage{epsfig}
\usepackage{graphicx}
\usepackage{dcolumn}
\usepackage{bm}
\usepackage{amsthm}
\usepackage{amsfonts}
\usepackage{color} 
\usepackage[usenames,dvipsnames]{xcolor}
\usepackage{amscd}
\usepackage{amsmath}
\usepackage{enumerate}
\usepackage{bbm}
\usepackage{epstopdf}
\usepackage[colorlinks=true,citecolor=blue,urlcolor=black]{hyperref}
\usepackage{subfigure}
\usepackage{enumitem}
\usepackage{color}
\usepackage{times}

\newtheorem*{theorem*}{Theorem}

\newcommand{\beq}{\begin{equation}}
\newcommand{\eeq}{\end{equation}}
\newcommand{\M}{\hat{\mathcal M}}
\newcommand{\N}{\hat{\mathcal N}}

\newcommand{\sT}{\mathsf{T}}

\newcommand{\rv}[1]{{\bf{#1}}}

\begin{document}
\title{Experimental fast quantum random number generation using high-dimensional entanglement \\ with entropy monitoring}

\author{Feihu Xu}
\email{fhxu@mit.edu}
\affiliation{%
Research Laboratory of Electronics, Massachusetts Institute of Technology, Cambridge, Massachusetts 02139, USA}

\author{Jeffrey H. Shapiro}
\affiliation{%
Research Laboratory of Electronics, Massachusetts Institute of Technology, Cambridge, Massachusetts 02139, USA}

\author{Franco N. C. Wong}
\affiliation{%
Research Laboratory of Electronics, Massachusetts Institute of Technology, Cambridge, Massachusetts 02139, USA}

\begin{abstract}
A quantum random number generator (QRNG) generates genuine randomness from the intrinsic probabilistic nature of quantum mechanics. The central problems for most QRNGs are estimating the entropy of the genuine randomness and producing such randomness at high rates. Here we propose and demonstrate a proof-of-concept QRNG that operates at a high rate of 24\,Mbit/s by means of a high-dimensional entanglement system, in which the user monitors the entropy in real time via the observation of a nonlocal quantum interference, without a detailed characterization of the devices. Our work provides an important approach to a robust QRNG with trusted but error-prone devices.
\end{abstract}

\maketitle

\section{Introduction}

Randomness is indispensable for a wide range of applications, ranging from Monte Carlo simulations to cryptography.  Quantum random number generators (QRNGs) can generate true randomness by exploiting the fundamental indeterminism of quantum mechanics~\cite{Ma15}. Most current QRNGs are photonic systems built with trusted and calibrated devices~\cite{trustQRNG1,trustQRNG2,trustQRNG3,trustQRNG4,trustQRNG5} that provide Gbit/s generation speeds at relatively low cost. However, a central issue for these QRNGs is how to certify and quantify the entropy of the genuine randomness, i.e., the randomness that originates from the intrinsic unpredictability of quantum-mechanical measurements. Entropy estimates for specific setups were recently proposed using sophisticated theoretical models~\cite{ma13,DF13,Haw13}. Nevertheless, these techniques require complicated device characterization that may be difficult to accurately assess in practice.

A solution to estimating the entropy is the device-independent (DI) or self-testing QRNG~\cite{C06,P10,C13}, but its practical implementation is challenging because it requires loophole-free violation of Bell's inequality, resulting in low generation rates of $\sim$1 bit/s~\cite{P10,C13}. Recently, Lunghi \emph{et al.} proposed a more practical solution that is based on a dimension witness~\cite{L15}, in which the randomness can be guaranteed based on a few general assumptions that do not require detailed device characterization. This scheme is highly desirable as it focuses on real-world implementations with trusted but error-prone devices, although its implementation with a two-dimensional (qubit) system still suffers from low generation rates of 10's of bits/s~\cite{L15}.

In this paper, we propose and experimentally demonstrate a fast QRNG operating at a rate of 24 Mbits/s, in which we can quantify and monitor, in real time, the entropy of genuine randomness without a detailed characterization of the trusted but error-prone devices. Our approach uses time-energy entangled photon pairs with high-dimensional temporal encoding~\cite{HDQKD}. High-dimensional temporal encoding is advantageous when the average interval between photon detection events is much longer than a time-bin duration set by the detector timing resolution. Such situations arise in typical quantum information processing tasks when single-photon detectors have long recovery times or when the pair generation rate must be kept low to minimize multi-pair events \cite{C13,L15}. The amount of genuine quantum randomness is quantified and monitored directly from observation of a nonlocal interference~\cite{Franson89}, and it is separated from other sources of randomness such as technical noise with a randomness extractor. We achieved the high generation rate by virtue of three experimental features: a high-dimensional time-energy entangled-photon source capable of producing multiple random bits per photon, a high-visibility Franson interferometer for evaluating entanglement, and high-efficiency superconducting nanowire single-photon detectors (SNSPDs). As a consequence, we are able to demonstrate a high-performance QRNG with a tolerance of device imperfections.

\section{Protocol} \label{sec:protocol}
Table~\ref{table:protocol} summarizes our protocol. An entanglement source generates high-dimensional entangled photon pairs. As an example, in the ideal case, the $N_d$-dimensional biphoton entangled state can be written as $|\psi\rangle = \frac{1}{\sqrt{N_d}}\sum^{N_d-1}_{i=0}|i\rangle_{\text{A}} \bigotimes |i\rangle_{\text{B}}\,,$
where $|i\rangle$ represents a single photon at a discretized time interval $i$. This state is observed by two measurement systems, one for random number generation (RNG) and the other for testing. Randomness is generated in the RNG mode from the state held by system A, $\rho_{\text{A}}$, which we assume is not pure and is correlated with environmental noise that models device imperfections. In the testing mode, the joint state is measured.

\begin{table}
\begin{center}
\begin{tabular}{p{0.95\linewidth}}\hline \hline 
\\
\includegraphics[width=0.9\columnwidth]{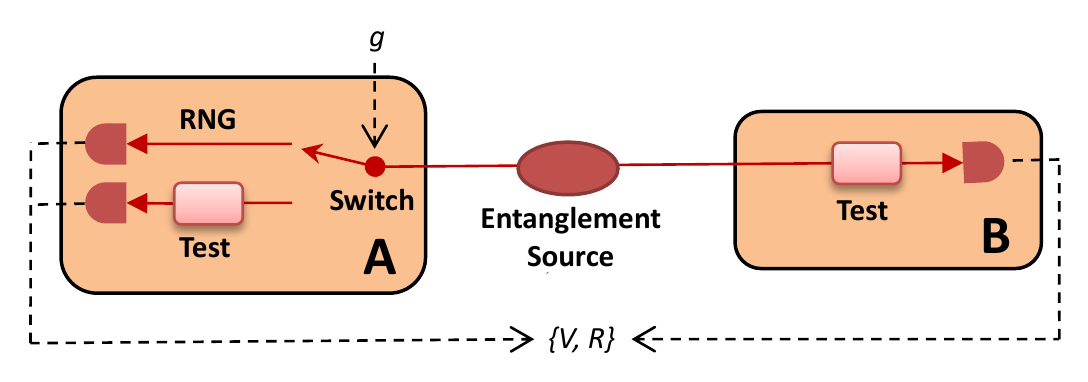}

\begin{enumerate}
\item There are $N$ measurement rounds. The entanglement source generates a high-dimensional entangled photon pair in each round. Bits $g_1, g_2, \ldots, g_N$ with values 0 or 1 are  independently chosen at random according to a $(q, 1-q)$ distribution.
\item For each round $i\in [N]$, if $g_i=0$, it is a generation round, then device $\rv{A}$ performs a time measurement in the `RNG' basis and outputs a random number $R_i$. If $g_i=1$,  it is a testing round, then devices $\rv{A}$ and $\rv{B}$ perform a joint frequency measurement in the `Test' basis.
\item From the results in the `Test' basis, calculate the testing value $V$. If $V$ exceeds a pre-set value $V_0$, then the protocol $\textbf{succeeds}$. Otherwise, it \textbf{aborts}.
\item If the protocol succeeds, the randomness throughput is evaluated from $V$ and a randomness extraction is applied to \{$R_i$\} to produce the genuine random numbers.
\end{enumerate} \\
\hline \hline
\end{tabular}
\end{center}
\caption{Protocol for quantum random number generation.} \label{table:protocol}
\end{table}

A key assumption of our approach is that the devices in the protocol are \emph{trusted}, namely they are not deliberately designed to fool the user, but the implementation may be imperfect. The central task is to estimate and monitor the amount of genuine randomness based \emph{only} on measurements. This is a nontrivial task as the observed randomness can have different origins. If the state $\rho_{\text{A}}$ is a superposition of high-dimensional states, then the outcome $R_i$ cannot be predicted with certainty, even if the internal state is known, thus resulting in genuine quantum randomness. On the other hand, the randomness may be due to technical imperfections such as detector noise and temperature fluctuations, whose randomness clearly has \emph{no} quantum origin, since the outcome $R_i$ can be perfectly guessed if the imperfections were well quantified.

In our approach, the amount of genuine randomness is monitored from the Franson visibility $V$~\cite{Franson89}. In particular, classical fields result in $V$ that is no greater than 50\%. For a maximally entangled state, $V$ would be 100\% in the ideal case~\cite{ZhongFranson}. Conceptually, $V>50\%$ guarantees that the source's output is entangled and thus contains genuine randomness. Rigorously, Ref.~\cite{ZZ14} has proven that, if we assume the biphoton wave function is Gaussian, $V$ provides an explicit bound for the correlations in frequency measurement, which in turn upper-bounds the conditional maximum entropy (given system $\rv{B}$) via the theory developed in~\cite{Furrer2012}. By using the entropic uncertainty relation for smooth entropies~\cite{vallone2014}, we can determine the conditional min-entropy given the environmental noise and thus the guessing probability, i.e., the amount of genuine randomness (see Appendix~\ref{App:sec1}).

Our approach is different from Ref.~\cite{L15} that is based on a dimension witness and was restricted to a two-dimensional system. Our system allows a much higher dimensionality that can be chosen in the post-processing step~\cite{HDQKD}, which was $N_d=2048$ dimensions in our experiment (see below). Our protocol provides self-monitoring because measurements of $V$ directly quantify the amount of genuine randomness in the observed data. A threshold value $V_0$ is pre-selected and the randomness can be generated \emph{only} when the observation satisfies $V>V_0$. Two particular advantages of this approach are: (i) the observation of $V$ does not rely on detailed models of the devices that are employed; and (ii) no loophole-free Bell inequality violation is required.

\section{Experiment}
\begin{figure}[tb]
\centering
\includegraphics[width=1.02\linewidth]{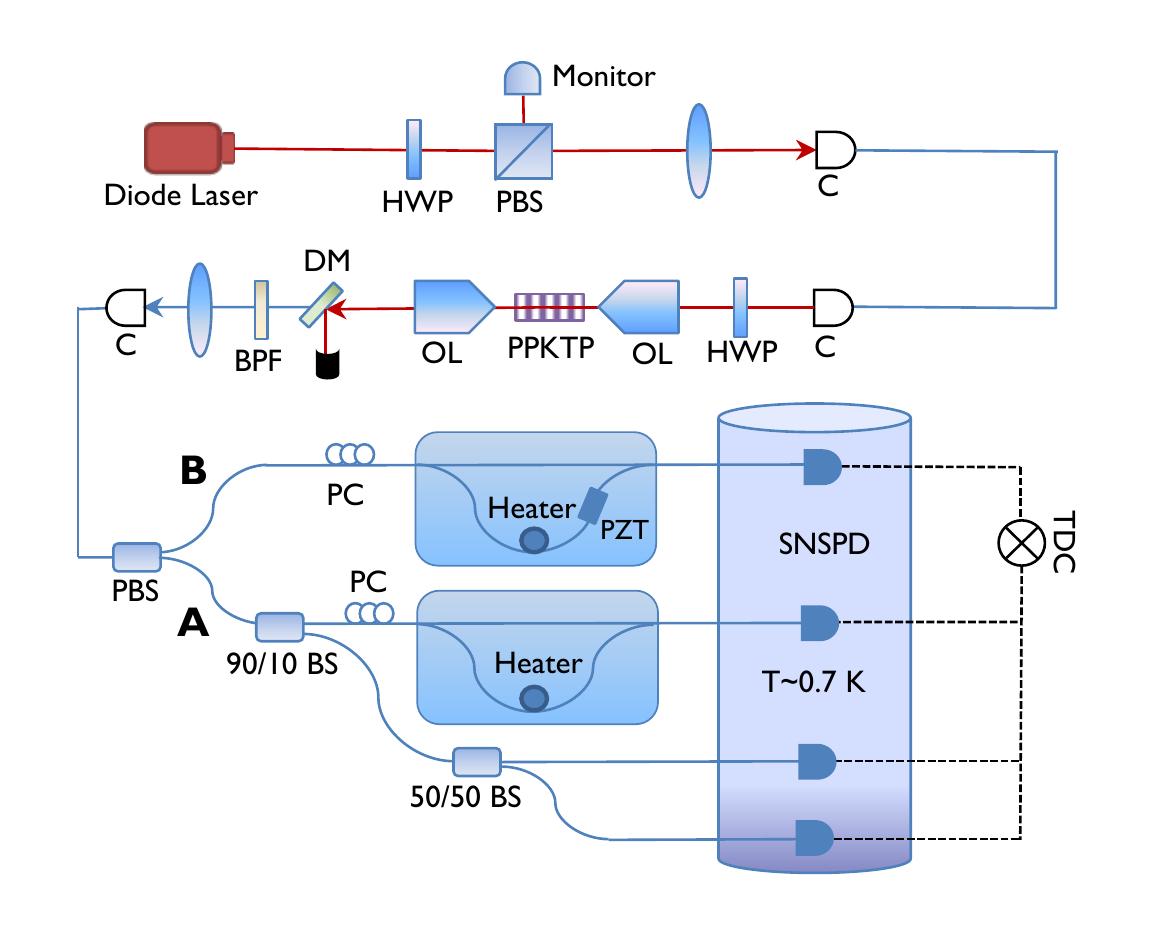}
\caption{Experimental setup. A cw diode laser pumps a PPKTP waveguide to generate time-energy entangled photon pairs. The orthogonally polarized signal and idler photons are coupled into a single-mode fiber, separated, and directed to $\rv{A}$ and $\rv{B}$, respectively. The signal photons in $\rv{A}$ are passively selected by a 90/10-ratio beam splitter for time-of-arrival measurement to generate random numbers or Franson measurement to check entanglement quality. HWP: half wave plate; PBS: polarization beam splitter; C: coupler; OL: objective lens; DM: dichroic mirror; BPF: band-pass filter; PC: polarization controller; BS: beam splitter; PZT: piezoelectric transducer; SNSPD: superconducting nanowire single-photon detector; TDC: time-to-digital converter.}
\label{Fig:setup}
\end{figure}

Figure~\ref{Fig:setup} shows the experimental setup. Time-energy entangled photon pairs were generated via spontaneous parametric down-conversion (SPDC) in a periodically-poled KTiOPO$_4$ (PPKTP) waveguide~\cite{Zhong12} that outputs multiple spatial modes at telecom wavelengths. The 46.1\,$\mu$m grating period was designed for type-II quasi-phase-matched wavelength-degenerate outputs at 1560\,nm in the fundamental modes of the signal and idler fields. The phase-matching bandwidth was 1.6\,nm with a corresponding biphoton correlation time of 2\,ps. The pump was a 780-nm continuous-wave (cw) diode laser with a measured coherence time of 2.2\,$\mu$s and the pump power coupled into the waveguide was monitored by a power meter. We extracted the fundamental signal and idler modes using a dichroic mirror to remove the pump and a 10-nm band-pass filter to spectrally remove the higher-order SPDC spatial modes. The fundamental modes were coupled into a standard single-mode fiber, achieving a $\sim$81\% waveguide-to-fiber coupling efficiency. We used a polarization beam splitter to separate the orthogonally polarized signal and idler photons and send them to devices $\rv{A}$ and $\rv{B}$, respectively. Losses in the waveguide and from the waveguide to the fiber were $\sim$15\% and $\sim$12\%, respectively. Overall, we measured a system efficiency of $\sim$50\% including the single-photon detector efficiency.

\begin{figure}[htbp]
\centering
\includegraphics[width=1\linewidth]{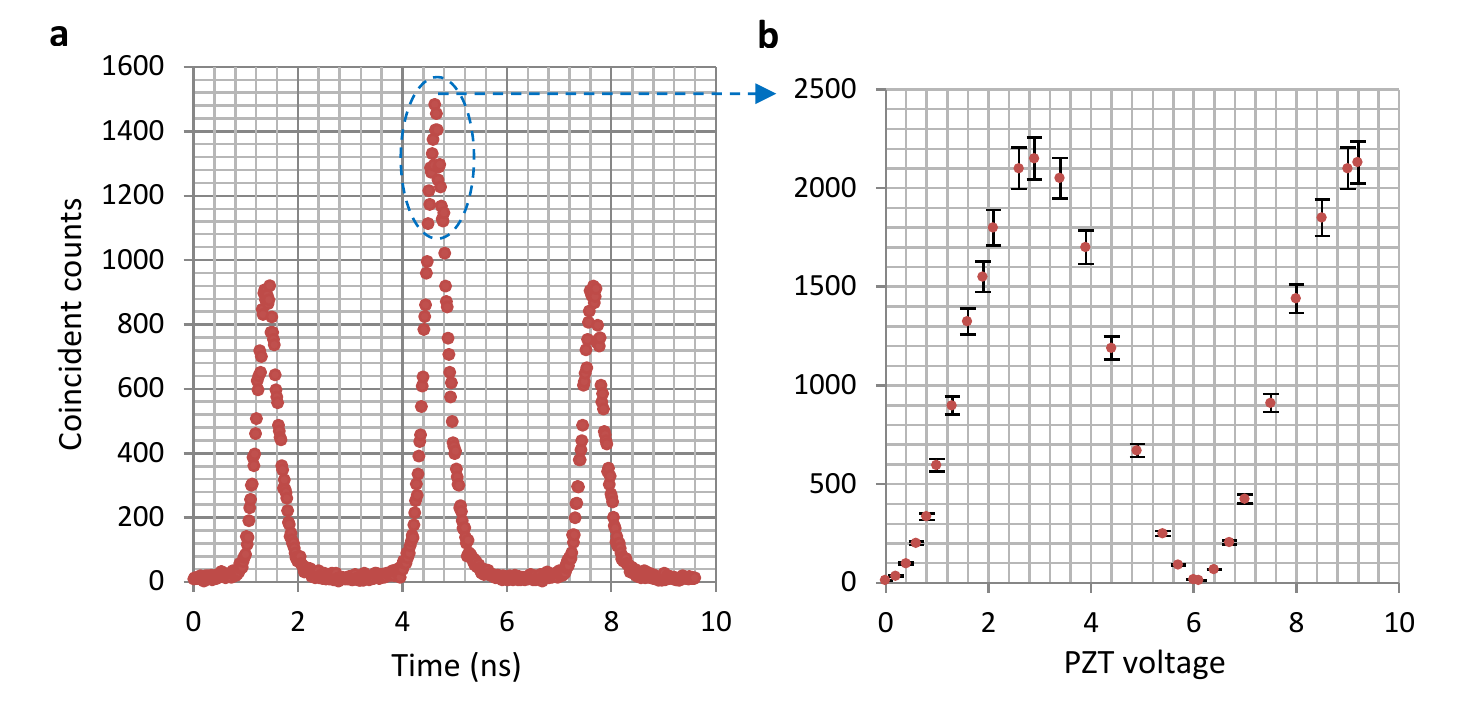}
\caption{Franson interference. (a) Typical distribution of signal and idler arrival-time difference in coincidence measurements. (b) Observed interference fringe at the central peak of (a) versus relative phase shift (proportional to PZT voltage). The observed raw visibility (without any subtraction) is $98.8\pm0.3\%$.}
\label{Fig:franson}
\end{figure}

A Franson interferometer is ideally suited for measuring the entanglement quality of a cw-pumped source of time-energy entangled light \cite{HDQKD,ZZ14}. We set up a Franson interferometer with  local dispersion cancellation that was comprised of two identical unbalanced Mach-Zehnder interferometers (MZIs), in which the long arm was made of standard single-mode fiber and low-dispersion LEAF fiber, such that the differential group delay (due to dispersion) between the long and short arms was zero~\cite{ZhongFranson}. To achieve long-term  stability, the MZIs were enclosed in a multilayered thermally insulated box, whose temperature was actively stabilized. The long-short path mismatch of each MZI was measured to be $\Delta T = 3.2$ ns. We coiled the long-path fiber of each MZI on a closed-loop temperature-controlled heater to precisely match the $\Delta T$ of the two MZIs. The variable relative phase shift between the two MZIs was set by a piezoelectric transducer fiber stretcher. By carefully fine-tuning the input polarizations and the temperatures, our time-energy entanglement source was found to have a Franson interference visibility $V$ of $98.8\pm0.3\%$, as shown in Fig.~\ref{Fig:franson}.

We performed a proof-of-concept implementation for the random basis choice, passively with a 90/10 beam splitter, i.e., $q = 0.9$ in Table~\ref{table:protocol}. The photon arrival times were measured by WSi SNSPDs~\cite{photonspot} that were placed in a closed-cycle cryogenic system with sub-Kelvin operating temperatures (see Fig.~\ref{Fig:setup}). The SNSPDs were measured to have detection efficiencies of $\sim$85\%, dark-count rates of $\sim$400/s, timing jitters of $\sim$250\,ps, and maximum count rates of $\sim$2 MHz without detector saturation. To mitigate the long reset times of the SNSPDs and to achieve a higher generation rate, system $\rv{A}$ used a passive 50/50 beam splitter to distribute incident photons equally between two WSi SNSPDs and their data were interleaved. Hence, a total of four WSi SNSPDs were used and their detection-time outputs were recorded by time-to-digital converters.

In the experiment, both the time-bin duration $\delta$ and the frame size (dimensionality) $N_d$ were chosen in the data post-processing step by parsing the raw timing records into the desired symbol length. As long as the frame duration $N_d\delta$ is smaller than the pump coherence time, we can precisely characterize the dimensionality. For experimental simplicity, we set $\delta$ equal to the detector timing jitter. $N_d$ was optimized to be 2048 in order to produce the maximal genuine randomness per photon (Appendix~\ref{App:sec1}).

\begin{figure}[htbp]
\centering
\includegraphics[width=0.9\linewidth]{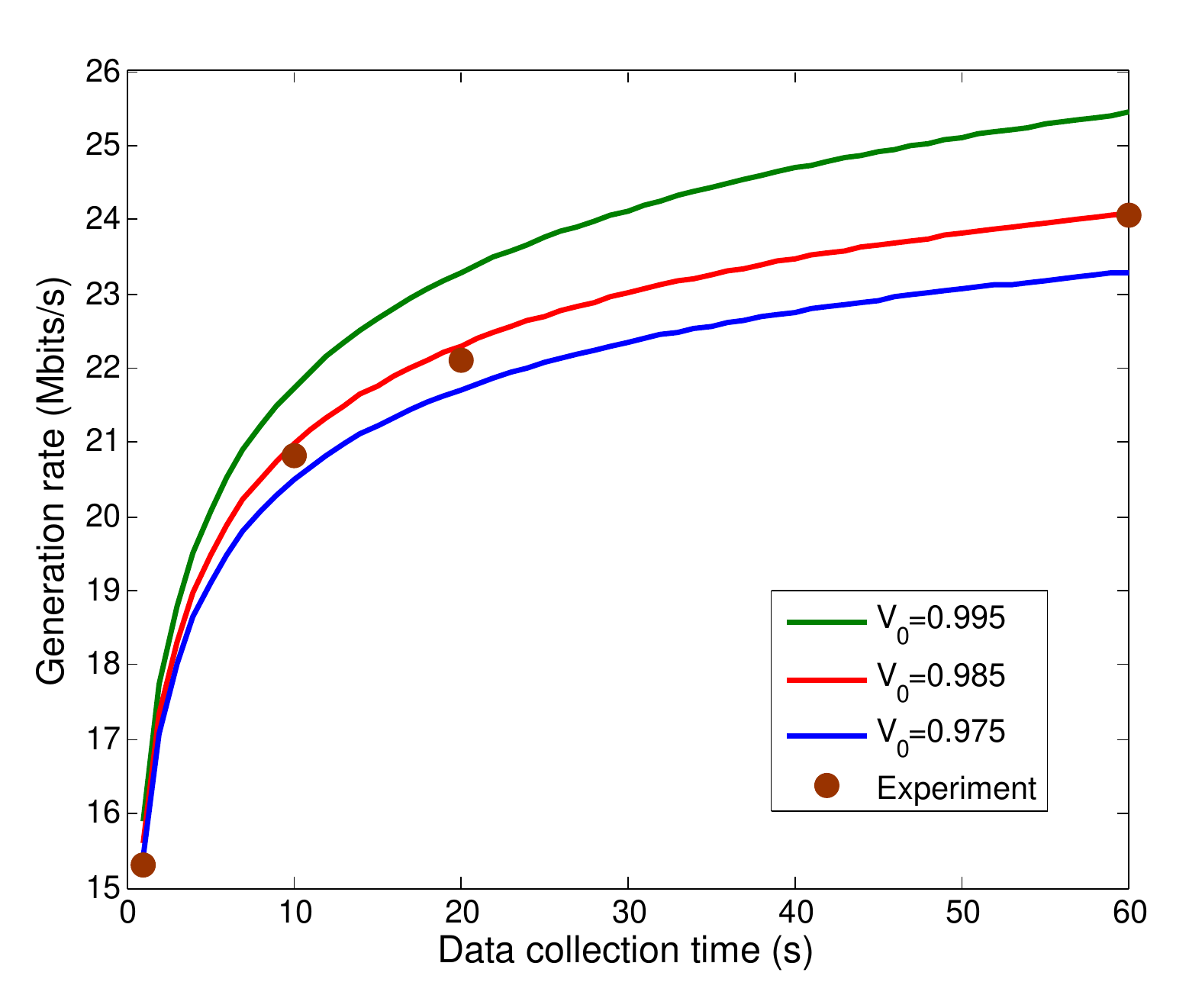}
\caption{QRNG throughput versus data collection running time. Solid red circles are experimental rates with $V_0=98.5\%$. The three curves are numerically evaluated results with different $V_0$ values, showing that a higher $V_0$ yields a higher throughput. For running times below 10\,s, the finite-data effect reduces the QRNG throughput substantially, whereas for a running time over 50 seconds, the throughput is already close to its asymptotic limit of 24.2 Mbit/s at $V_0 =98.5$\%.}
\label{Fig:output}
\end{figure}

In our proof-of-principle experiment, we recorded data for a maximum duration of 60\,s. We monitored the Franson visibility before and after data recording to ensure that the experimental $V$ exceeded the preset threshold $V_0$ in order to extract nonzero random bits from the data.  We set $V_0=98.5\%$ because it was the lower bound in most of the measurement runs in our experiment. We note that the amount of genuine randomness increases monotonically with increasing $V_0$, as shown in Fig.~\ref{Fig:output}, and therefore it is desirable to use high-quality devices in order to achieve high Franson visibility $V > V_0$.

We evaluated the auto-correlation of the raw generation-round data to be $\sim$0.001, which satisfies the independent, identically distributed (iid) assumption made in our analysis. Using the theory developed in Appendix~\ref{App:sec1}, we obtain 6.0 bits/photon genuine randomness at $N_d=2048$. After considering the unpaired-to-paired ratio of the SPDC output that we measured to be 1.8\% (see Appendix~\ref{App:sec2}), we extracted about 5.9 bits per $\log_2(2048)$-bit sample. We implemented a Toeplitz-hashing extractor~\cite{ma13} to extract genuine random numbers. A Toeplitz-hashing extractor extracts a random bit-string $m$ by multiplying the raw sequence $n$ with the Toeplitz matrix ($n$-by-$m$ matrix, random seed). The seed length of random bits required to construct the Toeplitz matrix is $d=n+m-1$. In our implementation with Matlab on a standard desktop computer, we chose the input and output bit-string lengths to be $n=4096$ and $m=4096 \times 5.9/11 \geq 2196$. Hence, a 4096-by-2196 Toeplitz matrix was generated in constructing the Toeplitz-hashing extractor. The output random bits successfully passed all the tests in the \textsc{diehard} test suite (see Appendix Table~\ref{Tab:Result:dihard}).

Figure~\ref{Fig:output} shows the experimental results for QRNG throughput for different running times. A longer running time produces more data, thus minimizing the finite-data effect and yielding more randomness from the raw data. The results show that a continuous running time of $\sim$60\,s can already produce randomness that is close to the asymptotic case of infinitely long operation. The measured count rates of the two SNSPDs for signal photon arrival times were 1.8 and 2.3 Mcounts/s. Hence, the final QRNG rate is $5.9\times 4.1= 24.2$ Mbit/s, which is many orders of magnitude higher than previous experiments based on self-testing~\cite{P10,C13} or a dimension witness~\cite{L15}. This dramatically faster rate benefits from the following factors: high-dimensional entanglement system that generates \emph{multiple} bits per photon, high-efficiency SNSPDs, high-quality PPKTP waveguide SPDC source with high fiber-coupling efficiency, and high-visibility Franson interferometry.

\section{Conclusion}
To sum up, we have demonstrated a QRNG based on high-dimensional entanglement with a rate over 24 Mbit/s. Compared to the standard device-dependent approaches with fully calibrated devices, our QRNG delivers a stronger form of security requiring less characterization of the physical implementation. The performance is close to commercial QRNGs~\cite{commercial}. Though our approach offers a weaker form of security than self-testing QRNG, it focuses on a scenario with trusted but error-prone devices. Together with other types of QRNGs demonstrated very recently in~\cite{sourceQRNG1,sourceQRNG2}, we believe that our results constitute an important step towards generating truly random numbers for practical applications.

Our proof-of-principle experiment can be further improved in the following directions. First, the Franson visibility was mainly limited by temperature fluctuations. Integrated photonics can improve the temperature stability, thus leading to higher interference visibility, as demonstrated recently in~\cite{IT16}. Second, the system can be operated at a different wavelength such as the visible or near-IR region, where inexpensive high-efficiency Si single-photon detectors can be used to replace the SNSPDs and potentially allow the system to be integrated with silicon photonics technology. Third, the randomness extraction was processed off-line by software, which can be improved with a field-programmable gate array implementation for real-time extraction. Fourth, the monitoring of the visibility can be done in real time by continuously observing the Franson measurement. Lastly, to increase the security of the QRNG protocol, we note that a loophole-free Bell's inequality test has been proposed for time-energy entanglement \cite{Cabello09}, thus making it possible to extend our high-dimensional entanglement system to function as a self-testing QRNG. Our QRNG is just one example of high-dimensional quantum information processing that we believe is an important area for future study and practical applications.

\section*{Funding Information}
Office of Naval Research (ONR) (N00014-13-1-0774); Air Force Office of Scientific Research (AFOSR) (FA9550-14-1-0052).

\section*{Acknowledgments}
The authors thank Murphy Yuezhen Niu, Tian Zhong, Xiaodi Wu, Zheshen Zhang for many helpful discussions.

\appendix

\section{Quantification of genuine randomness} \label{App:sec1}
The amount of genuine randomness is quantified from a pair of incompatible quantum measurements, namely time measurement $\mathbb T$ and frequency measurement $\mathbb W$. We take $\mathbb T$ and $\mathbb W$ to be positive-operator valued measures (POVMs) on $\rv{A}$ with elements $\{\M_t\}$ and $\{\N_w\}$, and random outcomes $T$ and $W$. Given an $N_d$-bin state observed by $\rv{A}$, the number of true random bits that can be extracted from $T$ that are independent of the environment system $\rv{E}$ is given by the conditional min-entropy $H_{\rm min}(T|\rv{E})$. Specifically, the probability of guessing $T$ by holding the system $\rv{E}$ is given by $p_{guess}(T|\rv{E})=2^{-H_\text{min}(T|\rv{E})}$~\cite{P10}. Hence, $H_{\rm min}(T|\rv{E})$ quantifies the genuine randomness. We bound $H_{\rm min}(T|\rv{E})$ based on the uncertainty relation~\cite{MT11,Furrer2012}, as proposed in~\cite{vallone2014}. While Ref.~\cite{vallone2014} demonstrated a QRNG with a maximal dimensionality of $N_d=4$, our system is capable for a much higher dimensionality, i.e., $N_d>2000$.

Consider three quantum systems $\rv{A}$, $\rv{B}$, and $\rv{E}$ and the tripartite state  $\rho_{ABE}$. The uncertainty relation can be written as~\cite{MT11}
\beq
\label{Eq:UP}
H_{\text{min}}(T|\rv{E})\geq -\log_2 c-H_{\text{max}}(W|\rv{B}),
\eeq
where $H_{\text{max}}(W|\rv{B})$ denotes the maximum entropy for $W$ given system $\rv{B}$, and $c$ is the maximum ``overlap'' between the two POVMs~\cite{MT11,Furrer2012}. By assuming that $\M_t$ and $\N_w$ are projective measurements corresponding to mutually-unbiased $N_d$-dimensional bases, then $ c=1/N_d$. Based on~\cite{Furrer2012}, we bound the maximum entropy as follows:
 \beq \label{Eq:uncertainty}
 H_{\text{max}}(W|\rv{B}) \leq \log_2 \gamma (d_w^{\text{L}_1}+\lambda),
 \eeq
where $d_w^{\text{L}_1}$ is the $\text{L}_1$ distance for correlations in the $\mathbb W$ basis,
\begin{align}\label{Eq:gamma}
\gamma(x)= (x+\sqrt{1+x^2})\Big(\frac{x}{\sqrt{1+x^2}-1}\Big)^{x},
\end{align}
and
\begin{align}\label{Eq:lambda}
\lambda \approx N_d \sqrt{\frac{1}{q(1-q)n_{\sT}}\ln\frac{1}{\epsilon_1/4-2f(p_{\alpha},n_{\sT})}} \,,
\end{align}
which quantifies the statistical fluctuations in the measurements.  In Eq.~(\ref{Eq:lambda}):
$n_{\sT}$ is the total number of detections in the $\mathbb T$ measurements; $f(p_{\alpha},n_{\sT})=\sqrt{2(1-(1-p_{\alpha})^{n_{\sT}})}$;  $p_{\alpha}$ is the probability of a biphoton's signal photon arriving outside a frame; and $\epsilon_1$ is the failure probability for the finite-data analysis, which was set to $\epsilon_1=10^{-10}$ in our experiment.

The $\text{L}_1$ distance $d_w^{\text{L}_1}$ can be bounded from the time-frequency uncertainty relation, as we now explain. The variances of the signal-idler time and frequency differences can be written as~\cite{ZZ14}
\begin{eqnarray} \label{Eq:Deltatw}
\langle(\Delta t)^2\rangle &=& \frac{e^{-2\beta}}{w_{0}^2}, \\ \nonumber
\langle(\Delta w)^2\rangle &=& e^{-2\beta}w_{0}^2\,.
\end{eqnarray}
Here $\beta$ is a squeezing parameter, $w_{0}$ is
\begin{align} \label{Eq:w0}
w_{0}=\frac{1}{\sqrt{2\sigma_{\text{coh}}\sigma_{\text{cor}}}}\,,
\end{align}
with $\sigma_{\text{coh}}$ being the pump coherence time and $\sigma_{\text{cor}}$ the biphoton correlation time.

We have assumed that the biphoton state has a Gaussian wave function. Reference~\cite{ZZ14} has proven that the variance of the frequency difference can be upper bounded from the Franson visibility via
\beq \label{Eq:Deltaw}
\langle(\Delta w)^2\rangle \leq \frac{2(1-V_0)}{\Delta T^2}\,.
 \eeq
Note that this bound applies only to high Franson visibilities, i.e., satisfying the assumptions made in~\cite{ZZ14}. By combining Eqs.~(\ref{Eq:Deltatw})--(\ref{Eq:Deltaw}), we arrive at the following upper bound of $d_w^{\text{L}_1}$
 \beq \label{Eq:dw1}
 d_w^{\text{L}_1}=\sqrt{\frac{2}{\pi}}\frac{\mid\langle(\Delta t)\rangle\mid}{\delta} \leq \frac{\sigma_{\text{coh}}\sigma_{\text{cor}}}{\delta\Delta T}\sqrt{\frac{16(1-V_0)}{\pi}}\,,
 \eeq
where $\delta$ is the time-bin duration selected in the protocol.

In the experiment, both the time-bin duration $\delta$ and the frame size (dimensionality) $N_d$ were chosen in the data post-processing step by parsing the raw timing records into the desired symbol length. $N_d$ was optimized to produce the maximal bits per photon: a larger $N_d$ can produce more raw bits per sample, i.e., $-\log_2 c$ in Eq.~\ref{Eq:UP}, but it also increases $H_{\text{max}}(W|\rv{B})$; hence, $N_d$ was optimized to maximize $H_{\text{min}}(T|\rv{E})$.

\section{Quantification of accidental counts} \label{App:sec2}
In the RNG round, the detections made by system $\rv{A}$ can be due to either SPDC signal photons or be accidental counts. Given our SNSPDs'  low dark-count rates, accidental counts are mainly due to the fluorescence (unpaired) photons in the SPDC's output. Here we quantify the SPDC output's unpaired-to-paired ratio.

The total number of photons/s $N_{\text{S}}$ generated in the signal field by SPDC can be written as a sum of paired photons/s $N_{\text{SPDC}}$ and fluorescence photons/s $N_{\text{F}}$:
\beq
N_{\text{S}}=N_{\text{SPDC}}+N_{\text{F}}.
 \eeq
Given the overall detection efficiencies for the signal ($\eta_{\text{S}}$) and idler ($\eta_{\text{I}}$) photons, the singles rate ($C_{\text{S}}$) and coincidence rate ($C_{\text{SI}}$) can be written as
\begin{eqnarray}
C_{\text{SI}} &=& N_{\text{SPDC}}\eta_{\text{S}}\eta_{\text{I}}, \\ \nonumber
C_{\text{S}} &=& N_{\text{S}}\eta_{\text{S}}.
\end{eqnarray}

In characterizing our entanglement source, a typical set of measurements yields the following values for the singles rate, coincidence rate, and efficiencies shown in Table~\ref{Tab:accidental}, and we obtain the unpaired-to-paired ratio for the signal field of the SPDC output
\beq
\frac{N_{\text{F}}}{N_{\text{SPDC}}}=1.8\%.
 \eeq
This result is consistent with reported ratios of 2\% in previous measurements of PPKTP waveguide and PPKTP bulk crystal at the telecom wavelengths~\cite{Zhong12}.

\begin{table}[hbt]
\begin{tabular}{c @{\hspace{0.5cm}} c @{\hspace{0.5cm}} c @{\hspace{0.5cm}} c}
\hline
\hline
$C_{\text{SI}}$ & $C_{\text{S}}$  & $\eta_{\text{S}}$ & $\eta_{\text{I}}$ \\
\hline
  420 kcoinc/s & 850 kcounts/s & 49.4\% & 50.3\% \\
\hline
\hline
\end{tabular}
\caption{Experimental values of singles rate, coincidence rate, and efficiencies. } \label{Tab:accidental}
\end{table}

\begin{table}[hbt]
\begin{tabular}{c @{\hspace{0.5cm}} c @{\hspace{0.5cm}} c}
\hline
\hline
Statistical test & $P$-value  & Result \\
\hline
  Birthday Spacings [KS]& 0.680563 & success\\
  Overlapping permutations & 0.308246 & success \\
  Ranks of 31x31 matrices & 0.450693 & success \\
  Ranks of 31x32 matrices & 0.591037 & success \\
  Ranks of 6x8 matrices [KS] & 0.448596 & success \\
  Bit stream test & 0.06551 & success \\
  Monkey test OPSO & 0.015300 & success \\
  Monkey test OQSO & 0.098700 & success \\
  Monkey test DNA & 0.098000 & success \\
  Count 1's in stream of bytes & 0.461867 & success \\
  Count 1's in specific bytes  & 0.031698 & success \\
  Parking lot test [KS] & 0.809513  & success \\
  Minimum distance test [KS] & 0.915470 & success \\
  Random spheres test [KS] & 0.902702 & success \\
  Squeeze test &  0.350940 & success \\
  Overlapping sums test [KS] & 0.795741  & success \\
  Runs test (up) [KS]& 0.569616 & success \\
  Runs test (down) [KS]& 0.248829 & success \\
  Craps test No. of wins & 0.259975 & success\\
  Craps test throws/game & 0.643893 & success \\
\hline
\hline
\end{tabular}
\caption{\textsc{diehard}. Data size is about 104 Mbits. For the cases of multiple $P$-values, a Kolmogorov-Smirnov (KS) test is used to obtain a final $P$-value, which measures the uniformity of the multiple $P$-values. The test is successful if all final $P$-values satisfy $0.01 \leq P \leq 0.99$.} \label{Tab:Result:dihard}
\end{table}

\end{document}